\documentstyle[sprocl,epsfig]{article}

\def\Journal#1#2#3#4{{#1} {\bf #2}, #3 (#4)}

\def\NPB{{\em Nucl. Phys.} B}
\def\PLB{{\em Phys. Lett.}  B}

\def\PRD{{\em Phys. Rev.} D}
\def\ZPC{{\em Z. Phys.} C}
\def\EJC{{\em Eur. Phys. J.} C}

\def\be{\begin{equation}}
\def\ee{\end{equation}}
\def\bea{\begin{eqnarray}}
\def\eea{\end{eqnarray}}

\def\myfigsize{0.66\linewidth}

\newcommand{\pom}{{I\!\!P}}

\begin{document}

\title{DIFFRACTIVE DIJET AND 3-JET ELECTROPRODUCTION AT HERA \\
(Talk presented at DIS 2000, Liverpool, April 2000)}

\author{F.-P. SCHILLING (for the H1 Collaboration)}

\address{Physikalisches Institut, University of Heidelberg, 
Heidelberg, Germany\\ e-mail: fpschill@mail.desy.de} 

\maketitle

\abstracts{ A new H1 measurement of diffractive dijet and 3-jet 
production cross sections in diffractive deep inelastic scattering
events of the type $ep\rightarrow eXY$ is presented. The data constrain
well the diffractive gluon distribution. At low $x_\pom$, a calculation
based on perturbative QCD is in a reasonable agreement with the data.
}

\section{Introduction}

\begin{figure}[!h]
\centering
\begin{minipage}{0.4\linewidth}
\centering
\epsfig{file=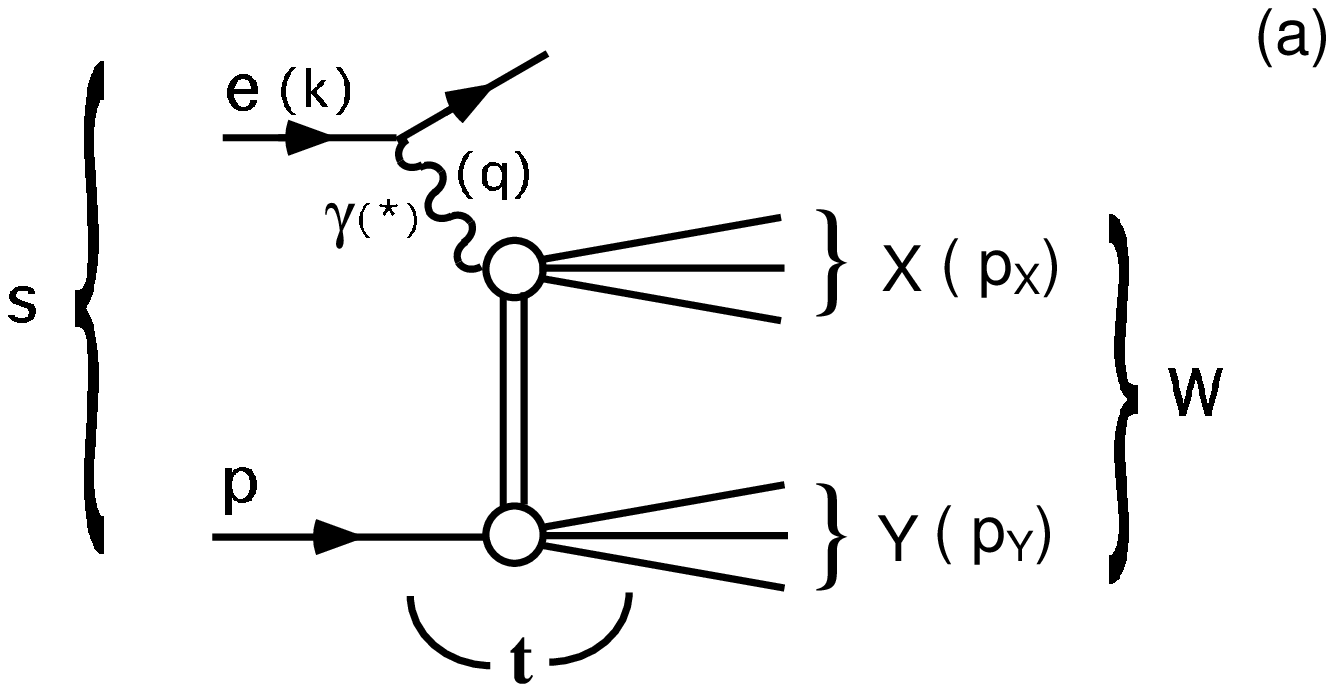,width=.8\linewidth}
\end{minipage}
\begin{minipage}{0.4\linewidth}
\centering
\epsfig{file=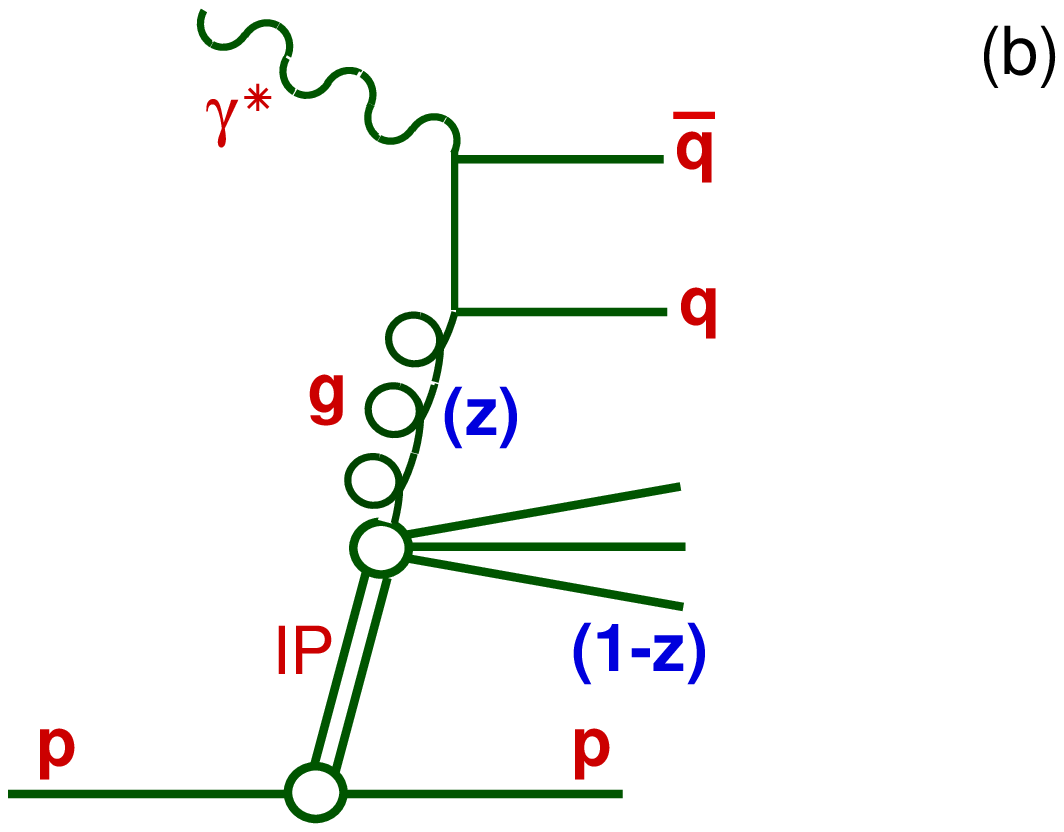,width=.4\linewidth}
\end{minipage}
\caption{(a) The generic diffractive process at HERA, where a photon from
the electron interacts with a proton via a net color singlet exchange,
producing final state hadronic systems $X$ and $Y$. Small $M_X$ 
and $M_Y$ are correlated with a large rapidity gap. 
(b) The dominating leading order QCD process for dijet production (BGF).
}
\label{diag}
\end{figure}

At the $ep$ collider HERA, color singlet exchange possessing vacuum quantum
numbers and traditionally assessed in terms of Regge phenomenology, can be 
studied using a virtual photon $\gamma^*$ as a probe (Fig.\ref{diag}a). 
High $p_T$ jet production is, in contrast to inclusive $F_2^{D}$ measurements,
directly sensitive to the gluonic structure of diffractive exchange 
(Fig.\ref{diag}b) and enables techniques of perturbative QCD to be applied.

Several approaches have been developed to describe the 
inclusive diffractive structure function $F_2^{D(3)}(x_\pom,\beta,Q^2)$ 
as measured at HERA, including
the resolved (partonic) pomeron model \cite{is}, Soft Color Interactions (SCI)
\cite{sci,sci2} or 
the Semiclassical model \cite{heb}.
Two-gluon exchange models usually take the proton rest frame point of 
view, where
the $\gamma^*$ is dissociating into a $q\overline{q}$ or $q\overline{q}g$ 
(dominant at large $M_X$, i.e. low $\beta$) state, scattering elastically 
off the proton by the exchange of a net color singlet pair of gluons.
A recent example is the Saturation model \cite{sat}, imposing a condition
of strong $k_T$ ordering on the gluon ($k_{T,g}\ll k_{T,q_i}$) in the case of
$q\overline{q}g$ production.
The high $p_T$ $q\overline{q}g$ configurations have
also been calculated without such a condition in perturbative QCD \cite{qqg}.

\begin{figure}[tb]
\centering
\epsfig{clip=,bbllx=0,bblly=283,bburx=567,bbury=520,file=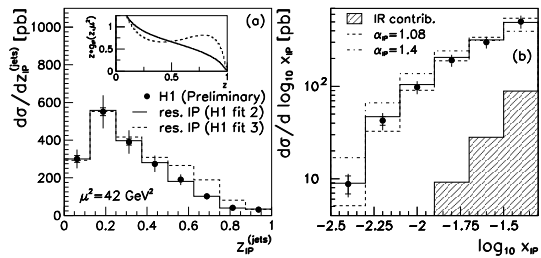,width=\myfigsize}
\caption{Diffractive dijet cross sections, compared to the 
resolved pomeron model.}
\label{fig1a}
\end{figure}

\begin{figure}[tb]
\centering
\epsfig{bbllx=0,bblly=283,bburx=567,bbury=520,file=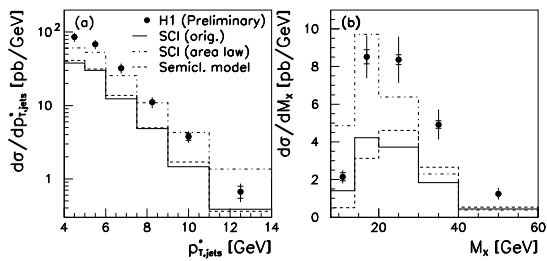,width=\myfigsize}
\caption{Dijet cross sections, compared to soft color neutralization models.}
\label{fig1b}
\end{figure}

\begin{figure}[tb]
\centering
\epsfig{bbllx=0,bblly=20,bburx=567,bbury=520,file=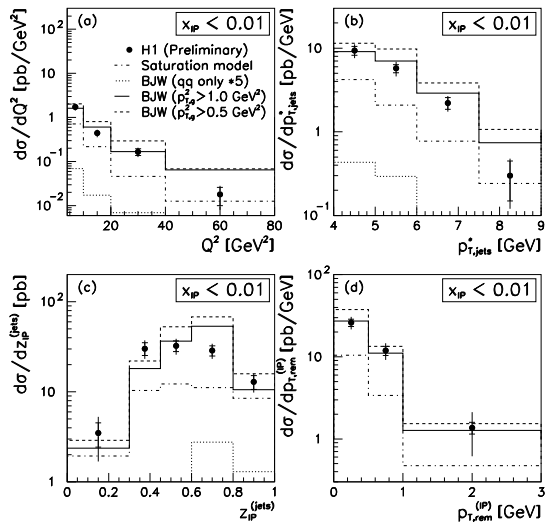,width=\myfigsize}
\caption{Dijet cross sections for $x_\pom<0.01$, compared the Saturation 
model and the calculations by Bartels et al. (``BJW'').
}
\label{fig2a}
\end{figure}
\begin{figure}[tb]
\centering
\epsfig{bbllx=0,bblly=283,bburx=567,bbury=520,file=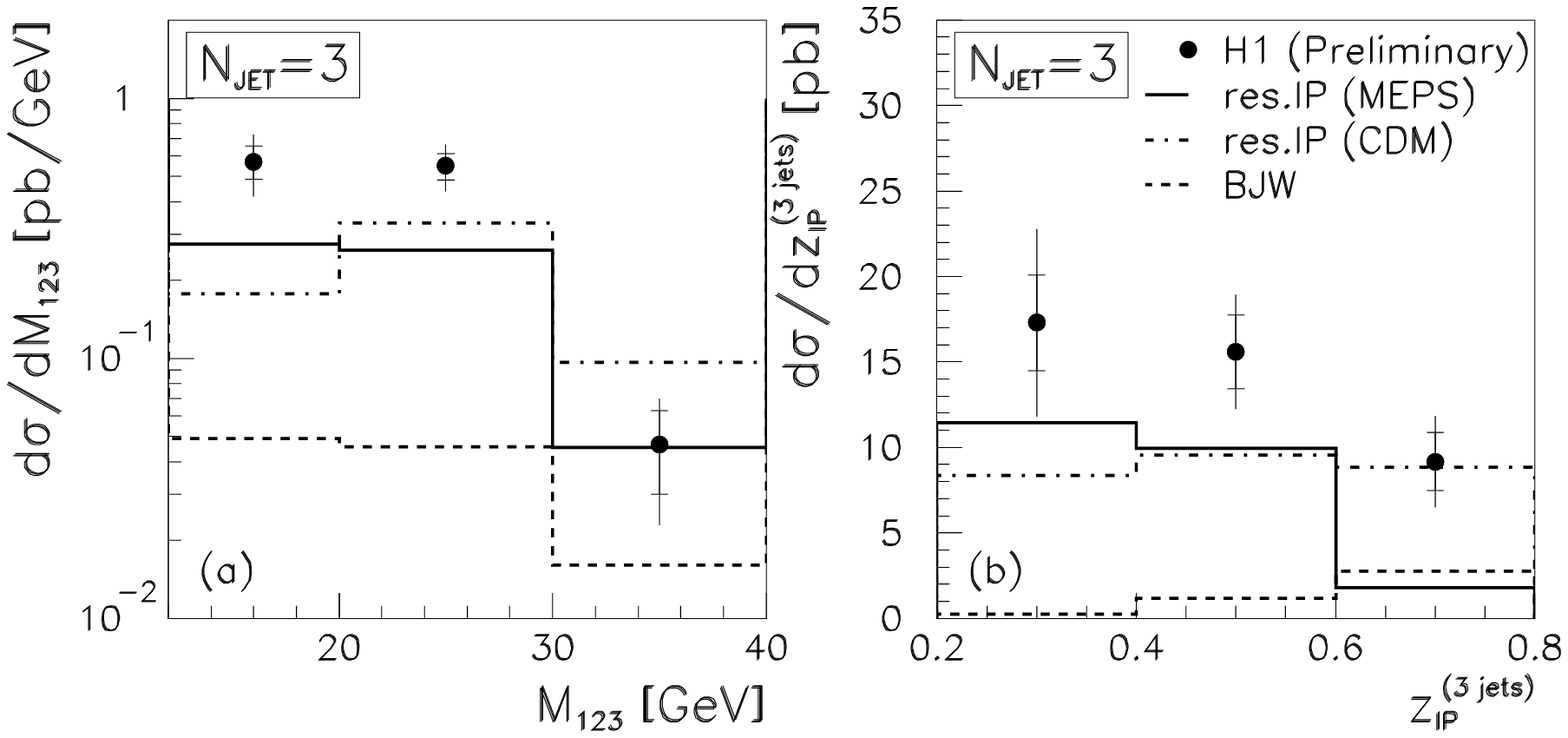,width=\myfigsize}
\caption{Diffractive 3-jet cross sections, compared to the resolved pomeron
model and the two-gluon model by Bartels et al.}
\label{fig2b}
\end{figure}

\section{Cross Section Measurement and Results} 

The presented cross sections have been extracted from 
$\mathcal{L} = \mathrm{18} \ \mathrm{pb}^{-1}$ of H1 data, using a rapidity 
gap selection and a cone jet algorithm with $R_{cone}=1.0$ in the $\gamma^* p$ 
frame. 
The selection yields approx. $2.500$ dijet and $130$ 3-jet events. 
The kinematic range of the measurement corresponds to 
$4 < Q^2 < 80 \rm\ GeV^2$, $x_\pom<0.05$, $|t|<1.0 \rm\ GeV^2$, 
$M_Y<1.6 \rm\ GeV$, $p_{T,jet}^*>4 \rm\ GeV$ and $-3 < \eta^*_{jet} < 0$\footnote{Quantities defined in the $\gamma^* p$ frame are labeled with a ``$^*$''}.
First results, based on a sub-sample of the data, were presented in 
\cite{myown}.
%

Fig.\ref{fig1a}a presents the measured dijet cross section as a function of 
$z_\pom=\beta(1+M^2_{12}/Q^2)$, corresponding within a partonic pomeron 
model to the pomeron momentum fraction entering the 
hard process. The prediction of the resolved (partonic) pomeron model, 
according to the QCD fits to $F_2^{D(3)}$ by H1 \cite{f2d3}, is also shown.
The data are well described if the ``fit 2'' (flat gluon) gluon density 
is used, whereas ``fit 3''
leads to an overestimate at high $z_\pom$. The corresponding gluon 
distributions are visualized in the insert. 
The dijet data constrain well the gluon distribution, in contrast to the
$F_2^D$ measurements. The $x_\pom$ cross section (Fig.\ref{fig1a}b)
visualizes a small Reggeon exchange contribution at high $x_\pom$.
The data are consistent with a pomeron intercept value of 
$\alpha_\pom(0)=1.2$, as 
obtained in \cite{f2d3}. Values of 1.08 (``soft pomeron'') or 1.4 
are  disfavored.

Fig.\ref{fig1b}, showing cross sections differentially in the mean jet
transverse momentum $p_{T,jets}^*$ and the mass of the $X$ system $M_X$, 
demonstrates that the original \cite{sci} and
the area-law-improved \cite{sci2} versions of SCI and the Semiclassical model
fail either in shape or normalization to describe the data. However,
NLO contributions have not yet been taken into account.

Fig.\ref{fig2a} presents dijet cross sections for the restricted
kinematic region of $x_\pom<0.01$, avoiding the valence quark 
region in the 
proton and secondary exchange contributions. The data are compared to the 
Saturation model and the calculations of Bartels et al.
The Saturation model, which imposes the condition $k_{T,g}\ll k_{T,q_i}$,
underestimates the cross section. 
Within the BJW model, the contribution of $q\overline{q}$
states alone is negligible in the covered region of small $\beta$. If 
the $p_T$-cutoff for the gluon in the case of
$q\overline{q}g$ production is set to $p^2_{T,g}>1.0 \rm\ GeV^2$, a rough
agreement
with the data is achieved with only one additional free parameter. 
Lowering this cutoff leads to an overestimate of the cross section,
becoming visible esp. in Fig.\ref{fig2a}d, the $p_T$ distribution of the
hadronic final state not belonging to the two jets in the $\pom$ hemisphere.

In Fig.\ref{fig2b}, the measured 3-jet cross sections are presented
as functions of the 3-jet invariant mass $M_{123}$ and
$z_\pom^{(3 \ jets)}$, a $z$-variable defined for 3 jets. 
The partonic pomeron model, incorporating
two approximations for QCD diagrams beyond leading order,
the parton shower (MEPS) and color dipole (CDM) models, is below the 
data. The BJW two-gluon exchange calculation, well suited for 3-jet production
in principle, yields too small cross sections in this phase space region, 
which is kinematically bound to the region of large $x_\pom$.

\section{Conclusions} 

The measurement of diffractive jet production is complementary to $F_2^{D(3)}$
measurements because it constrains well the diffractive gluon distribution.
It is powerful in discriminating between different models for 
diffraction and can isolate perturbatively treatable 
contributions to $\sigma_{diffr.}$ at low $x_\pom$, where a reasonable 
agreement with a pQCD 2-gluon model is found.

\end{document}